\documentclass[twocolumn,fleqn]{article} \usepackage{amsmath,amssymb,amsfonts}
\usepackage{graphicx} 			% for TeX 2016 version of arXiv
\usepackage{url}
\usepackage{siunitx} 
\title {Investigating factors behind the outbreak of the 6th and the 7th waves of COVID-19 in Tokyo} 
\author {Yoshihiko Takase \thanks {Chiba JICA Senior Volunteers Association}} 
\begin{document} 
\maketitle 

% --- Abstract --- 
\abstract {}
The 6th wave of COVID-19 in Tokyo continued for the longest period of infection (about 190 days from late Nov. 2021), and the 7th wave, which occurred in mid-May 2022, was the largest wave ever (cumulative 1.7 million people). In order to elucidate their factors, the infection wave was analyzed by using the Avrami equation.
The main component of the 6th wave was formed by the coupling of increased human interaction due to the New Year holidays and the invasion of the new virus variant Omicron BA.1. After that, side waves were formed by the coupling of the invasion of the new virus variant Omicron BA.2 and the human interaction in the consecutive holidays in February, March, and May. These side waves caused the 6th wave not to converge for a long time.
The outbreak of the main component of the 7th wave occurred by the coupling of the invasion of the new virus variant Omicron BA.5 and multiple social factors, followed by human interaction during the July holidays.
Based on the results that the domain growth rate $K$ and the infection rise time $t_\mathrm{on}$ were almost independent of the initial susceptible $D_\mathrm{s}$, dense nucleation followed by a near growth model was deduced.
The quantity $K \cdot t_\mathrm{on}$ was considered to represent the infectivity of the virus.

% --- Introduction --- 
\section {Introduction}

The 6th wave of COVID-19 in Tokyo continued for the longest period of infection (about 190 days from late Nov. 2021), and the 7th wave, which occurred in mid-May 2022, was the largest wave ever (approximately 1.7 million people in total)\cite{metro_tokyo}. 

We have regarded the spread of COVID-19 infection as a phase transition and been analyzing it using the Avrami (or JMAK) equation\cite{avrami,mandelkern}. The advantage is that a straightforward model of nucleation and growth can be assumed, the model can be described by a single equation\cite{mandelkern}, and then it is relatively easy to apply the least-squares method to determine the optimum values of the parameters\cite{ferro_pol_rev,nylon_pol_rev}.

Based on the random nucleation and subsequent linear domain-growth model, the 1st to the 5th waves of COVID-19 in Japan were analyzed and the parameters were determined\cite{sim_covid19_1_japan,sim_covid19_1to5_japan,covid19_tokyo}.
The accuracy of the simulated cumulative infection $D$ was as good as $95\% CI / D < 2.5\%$ after the 2nd wave when $D$ exceeded 10,000\cite{sim_covid19_1to5_japan}. 

Since this method is highly accurate, the factors behind the infection may be elucidated by simulating the infection waveform in detail.
This time, we will focus on the infection wave in Tokyo, the capital of Japan, where various data have been published. 

The purpose of this study is to simulate the  detailed structure of the infection waveform in Tokyo by using the Avrami equation and to elucidate the factors behind it.

% --- Waveform analysis --- 
\section {Waveform analysis of the COVID-19 infection in Tokyo} 
\subsection {Simulation characteristics of the 5th to the 7th waves}

The 5th, 6th, and 7th waves in Tokyo were simulated by superposition of 5 waves (5.1th to 5.5th waves), 6 waves (6.1th to 6.6th waves), and 5 waves (7.1th to 7.5th waves), respectively. 

\begin {figure} [p] 
	\centering 
	\includegraphics[width=8.0cm]{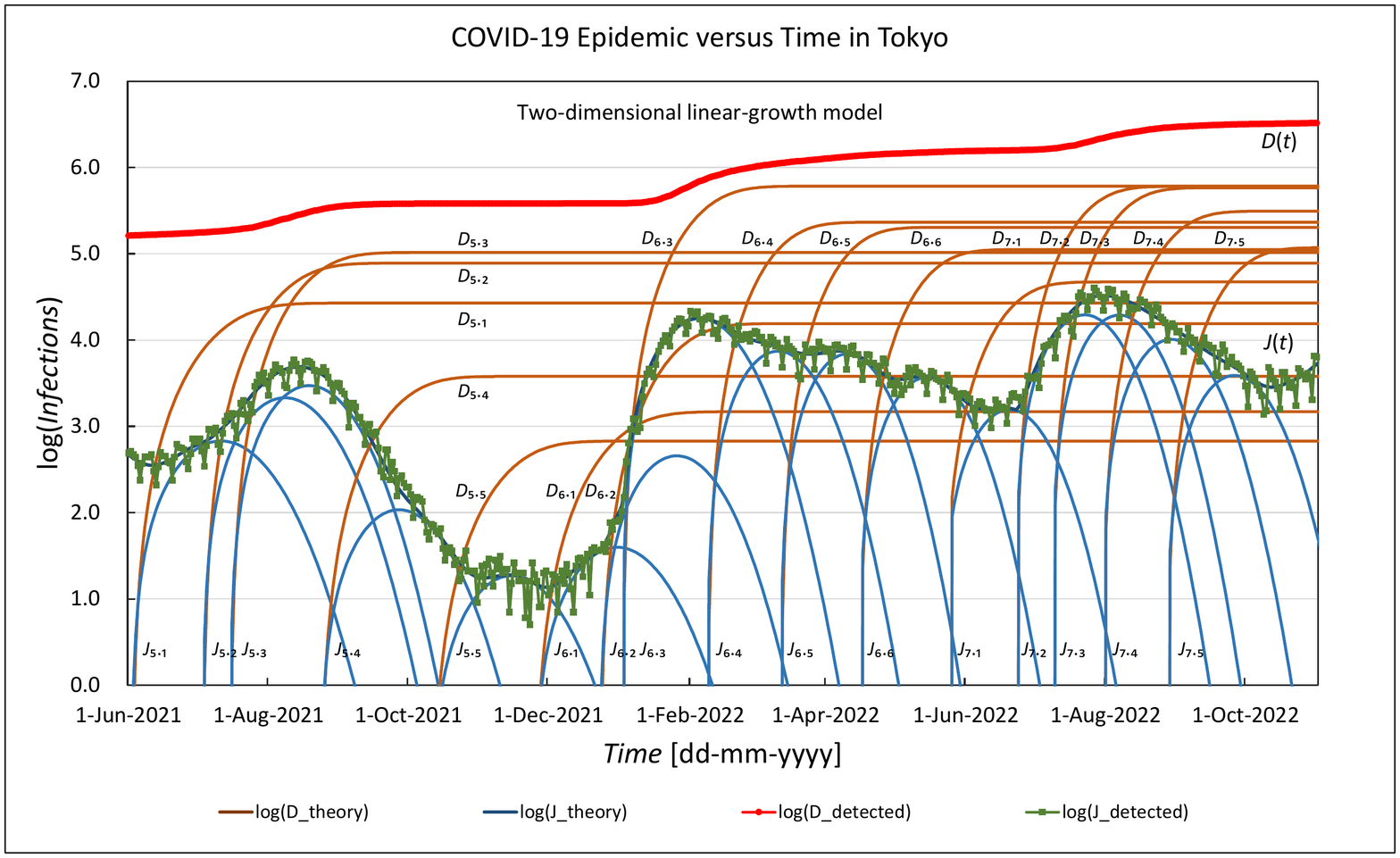} 
	\caption {The semi-logarithmic representation of the detected and the simulated values of the daily infections $J(t)$ and the total infections $D(t)$ for the 5th, the 6th, and the 7th waves in Tokyo.
	}
	\label {fig:allwaves_semilog} 
\end {figure}

Since the magnitude of each wave differs greatly, the number of time-dependent daily new infections $J(t)$ (hereinafter referred to as the daily infections) and  the number of time-dependent cumulative daily new infections $D(t)$ (hereinafter referred to as the total infections) for the 5th to the 7th waves (June 1, 2021 to November 1, 2022) are shown in semi-logarithmic representation in Fig. \ref{fig:allwaves_semilog}, where $t$ is the time. 
The dark green and the red markers are the detected daily and total infections, respectively. The blue curve is the theoretical value of the daily infections for each wave and the dark blue curve is the sum of them, $J(t)$. The brown curve is the theoretical value of the total infections for each wave and the dark brown curve is the sum of them, $D(t)$. 
The accuracy of the simulated total infections $D(t)$ was as good as $95\% CI / D(t) < 0.8\%$ after the 5th wave.

Start day and end day of the duration of each wave to which the least-squares method is applied to find the parameters for the nucleation and subsequent two-dimensional domain growth model is shown in Table \ref{table:waves_duration}. These durations were determined by iteratively applying the least-squares method to improve simulation accuracy\cite{sim_covid19_1to5_japan}.

% --- Table of the duration of each wave to which the least-squares method is applied --- 
\begin{table} %[h] 
\caption{Start day and end day of the duration of each wave to which the least-squares method is applied and the durations of the 5th to the 7th waves.\\} 
	\label{table:waves_duration} 
	\centering
	\begin{tabular} {cccc} 
	\hline 
	Wave & Start day & End day & Duration\\ 
	  &   &   & [day] \\ 
	\hline \hline 
	5.1th & 02-Jun-2021 & 04-Jul-2021 & \\ 
	5.2th & 03-Jul-2021 & 22-Jul-2021 & \\ 
	5.3th & 15-Jul-2021 & 26-Sep-2021 & \\ 
	5.4th & 23-Aug-2021 & 24-Oct-2021 & \\ 
	5.5th & 10-Oct-2021 & 01-Dec-2021 & 182 \\ 
	\hline 
	6.1th & 24-Nov-2021 & 26-Dec-2021 & \\ 
	6.2th & 23-Dec-2021 & 03-Jan-2022 & \\ 
	6.3th & 02-Jan-2022 & 21-Feb-2022 & \\ 
	6.4th & 08-Feb-2022 & 22-Mar-2022 & \\ 
	6.5th & 12-Mar-2022 & 24-Apr-2022 & \\ 
	6.6th & 15-Apr-2022 & 29-May-2022 & 186 \\ 
	\hline 
	7.1th & 21-May-2022 & 20-Jun-2022 & \\ 
	7.2th & 22-Jun-2022 & 18-Jul-2022 & \\ 
	7.3th & 08-Jul-2022 & 11-Aug-2022 & \\ 
	7.4th & 30-Jul-2022 & 11-Sep-2022 & \\ 
	7.5th & 27-Aug-2022 & 10-Oct-2022 & 142 \\  
	\hline 
	\end{tabular} 
\end{table}

% --- subsection --- 
\subsection {Determined parameters for each wave} 

The parameters determined for each wave, $D_\mathrm{s}$, $K^2$, $K$, and $t_\mathrm{on}$ are shown in Table \ref{table:three_parameters}, where $D_\mathrm{s}$ is the initial susceptible, $K$ the domain growth rate, and $t_\mathrm{on}$ the rise time defined as the time for the $D(t) - t$ characteristic to change from its 10\% value to its 90\% value, originally an electronics definition. The value of the nucleation decay constant $\nu$ was assumed to be constant ($\nu = 0.0090$ [1/day]) for the reason described in the previous report\cite{covid19_tokyo}.

% --- Table of three paremeters --- 
\begin{table} %[h] 
  \caption {Parameters determined for each wave, $D_\mathrm{s}$, $K^2$, $K$, and $t_\mathrm{on}$.\\} 
  \label {table:three_parameters}
  \centering 
  \begin {tabular} {crrrr} 
	\hline 
	Wave & \multicolumn{1}{c}{$D_\mathrm{s}$} & \multicolumn{1}{c}{$K^2$} & \multicolumn{1}{c}{$K$} & \multicolumn{1}{c}{$t_\mathrm{on}$}\\ 
	& [person] & [$1/\si{day}^2$] & \multicolumn{1}{c}{[$1/\si{day}$]} & \multicolumn{1}{c}{[$\si{day}$]} \\ 
	\hline \hline 
	5.1th & 26,825 & 0.00780 & 0.0883 & 39 \\ 
	5.2th & 78,015 & 0.00996 & 0.0998 & 36 \\ 
	5.3th & 103,593 & 0.01119 & 0.1058 & 35 \\
	5.4th & 3,791 & 0.01102 & 0.1050 & 35 \\
	5.5th & 674 & 0.01005 & 0.1003 & 36 \\
	\hline 
	6.1th & 1,477 & 0.00933 & 0.0966 & 38 \\ 
	6.2th & 15,520 & 0.01198 & 0.1095 & 34 \\ 
	6.3th & 607,980 & 0.01182 & 0.1087 & 34 \\
	6.4th & 232,055 & 0.01546 & 0.1243 & 31 \\
	6.5th & 202,709 & 0.01840 & 0.1356 & 30 \\
	6.6th & 111,942 & 0.01693 & 0.1301 & 30 \\
	\hline 
	7.1th & 47,148 & 0.01710 & 0.1308 & 30 \\ 
	7.2th & 604,586 & 0.01600 & 0.1265 & 31 \\ 
	7.3th & 582,825 & 0.01740 & 0.1319 & 30 \\ 
	7.4th & 311,964 & 0.01626 & 0.1275 & 31 \\ 
	7.5th & 117,535 & 0.01695 & 0.1302 & 30 \\ 
	\hline
  \end {tabular} 
\end {table}

% --- Discussions --- 
\section{Waveform analysis of the COVID-19 6th infection in Tokyo} 
% --- subsection The 6th wave in Tokyo --- 
\subsection{Simulation characteristics of the 6th wave} 

\begin {figure} %[p] 
	\centering 
	\includegraphics [width=8.0cm] {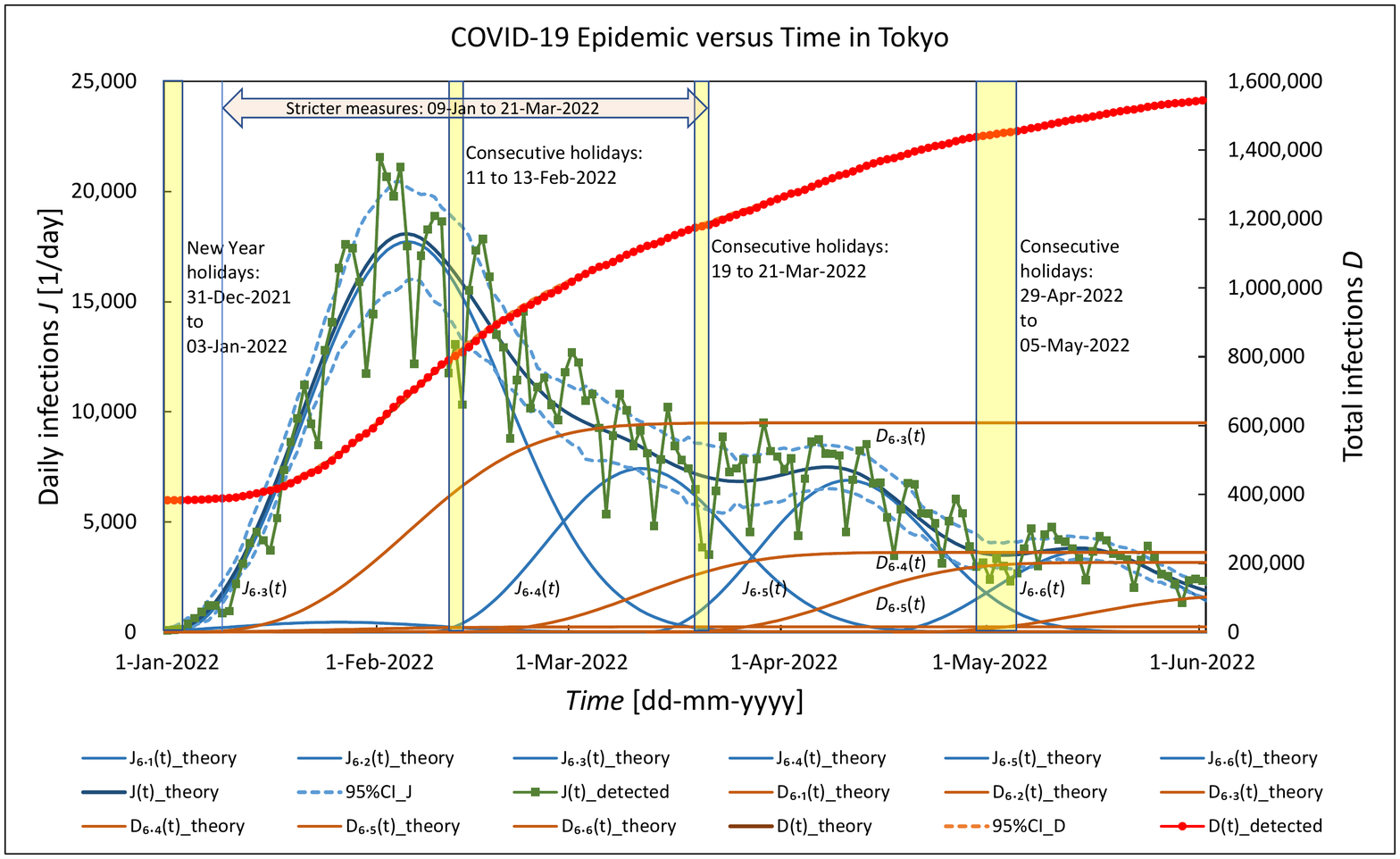}
	\caption {The detected and the simulated values of the daily infections $J(t)$ and the total infections $D(t)$ of the 6th wave in Tokyo from January 1, 2022 to June 1, 2022. Symbols: refer to Fig. \ref{fig:allwaves_semilog}. The broken line is the $95\% CI$.
	}
	\label {fig:wave6} 
\end {figure}

Figure \ref{fig:wave6} shows the time-dependent characteristics of the detected and the simulated values of the daily infections $J(t)$ and the total infections $D(t)$ for the 6th wave in Tokyo from January 1, 2022 to June 1, 2022. The meaning of each marker and line is the same as that in Fig. \ref{fig:allwaves_semilog}. The broken line is the $95\% CI$ calculated from the moving standard deviation with 7-days sliding window.

% --- subsection --- 
\subsection{The correlation between the 6th wave and the BA.1 variant ratio}

\begin {figure} %[p] 
	\centering 
	\includegraphics [width=8.0cm] {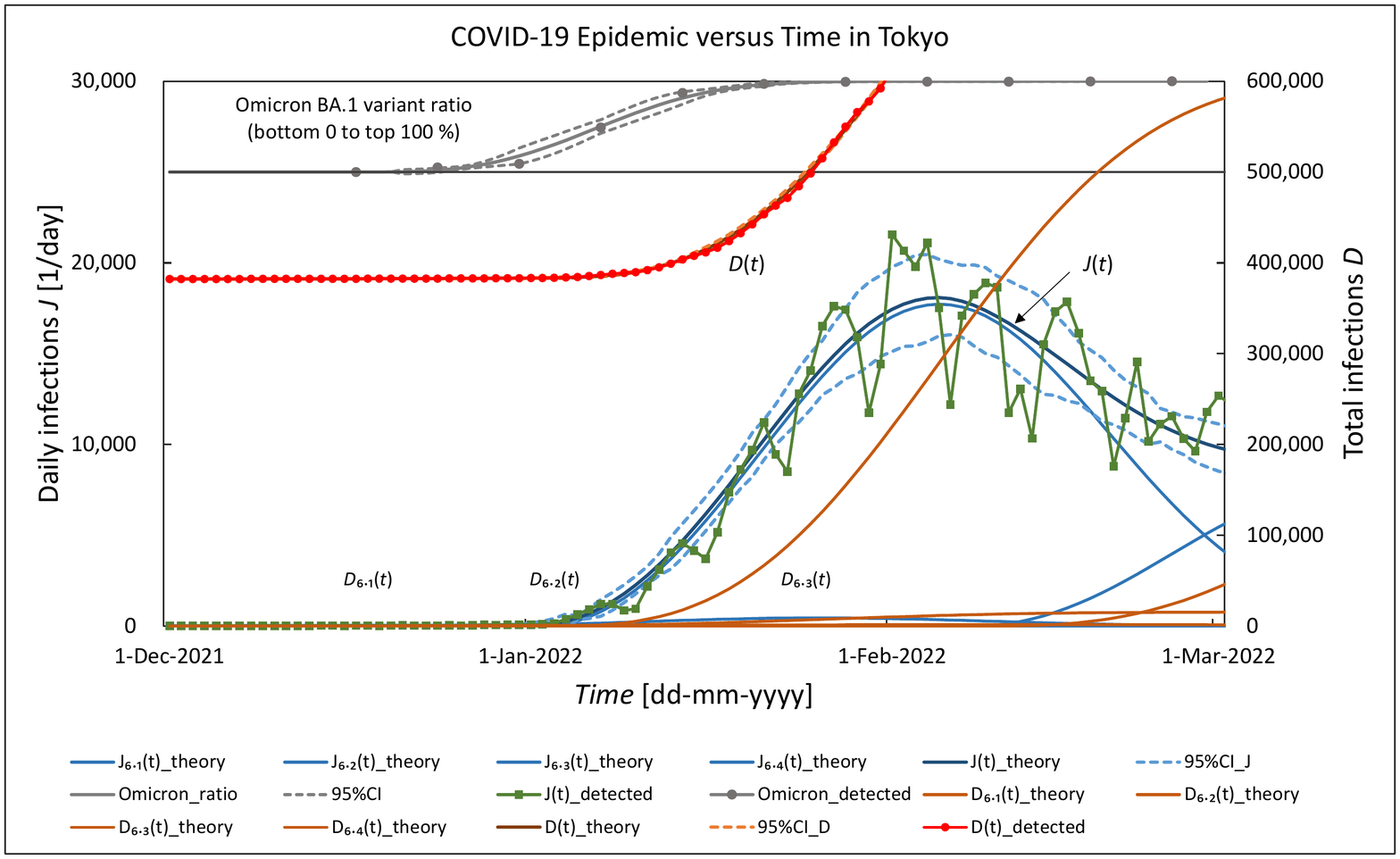}
	\caption {The characteristics of the 6th wave with the data of $p_\mathrm{BA1}(t)$. Symbols: refer to Fig. \ref{fig:wave6}.
	}
	\label {fig:inf_ba1_wave6} 
\end {figure}

Shortly before the start of the 6th wave, Delta variant of the virus was replaced by the new virus variant, Omicron BA.1\cite{voc_who}. Figure \ref{fig:inf_ba1_wave6} shows the characteristics of the 6th wave with the percentage $p_\mathrm{BA1}(t)$ of the suspected Omicron BA.1 variant from the (84th) Tokyo Metropolitan Government Novel Coronavirus Infectious Diseases Monitoring Meeting materials\cite{monitoring84} inserted in the upper part. The value of $p_\mathrm{BA1}(t)$ is also well approximated by the Avrami equation. The 6.1th and the 6.2th waves are too small to be seen in Fig. \ref{fig:inf_ba1_wave6}, but can be seen in Fig. \ref{fig:allwaves_semilog}.

\begin {figure} %[p] 
	\centering 
	\includegraphics [width=8.0cm] {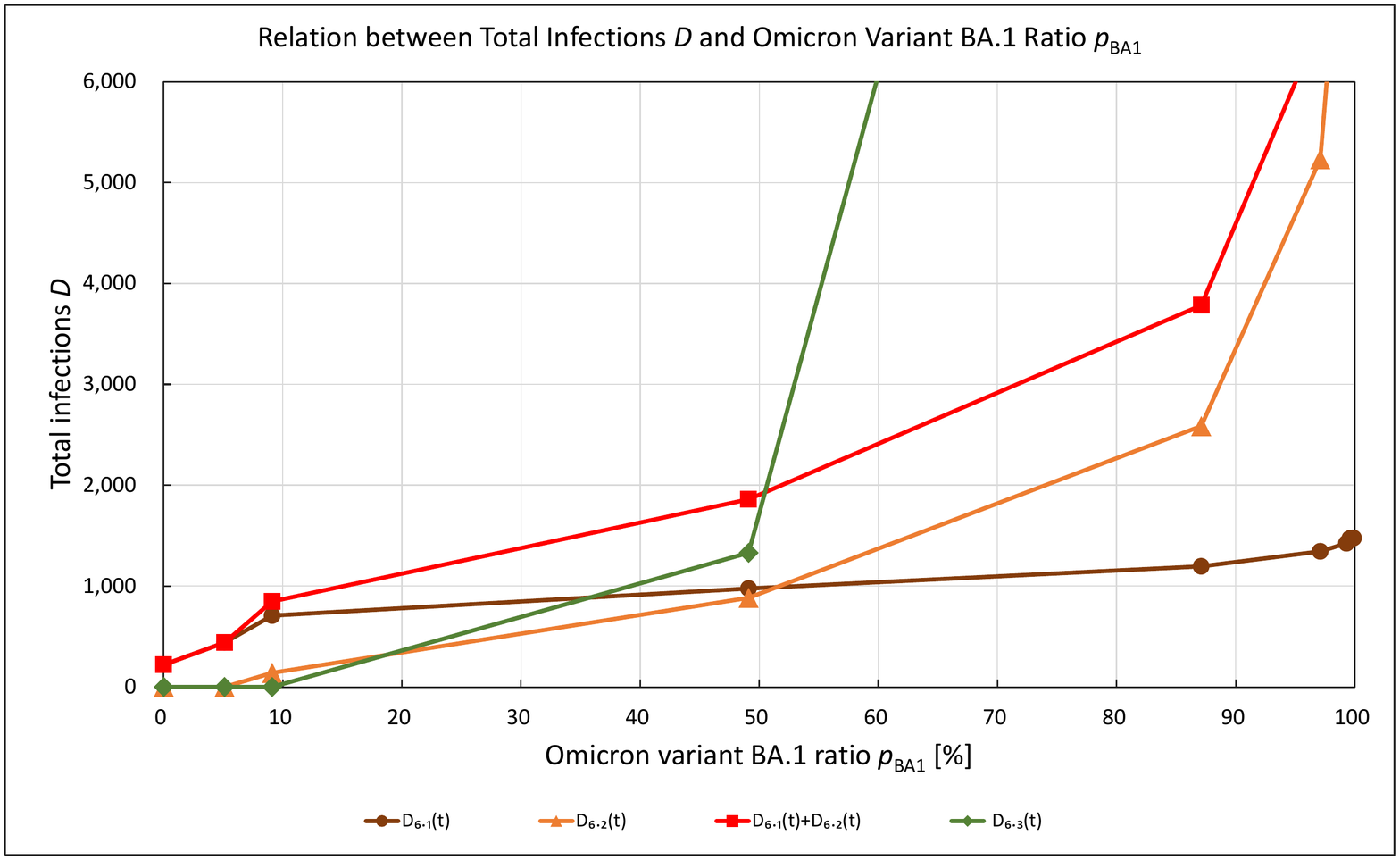}
	\caption {The relationship between $D_\mathrm{6.1}, D_\mathrm{6.2}, D_\mathrm{6.1}+D_\mathrm{6.2}, D_\mathrm{6.3}$ and $p_\mathrm{BA1}$. 
	}
	\label {fig:corr_inf_ba1_wave6} 
\end {figure}

Figure \ref{fig:corr_inf_ba1_wave6} shows the relationship between $D_\mathrm{6.1}, D_\mathrm{6.2}, D_\mathrm{6.1}+D_\mathrm{6.2}, D_\mathrm{6.3}$ and $p_\mathrm{BA1}$ to quantitatively understand the virus mutation effect. The relationship between $D_\mathrm{6.1}$ and $p_\mathrm{BA1}$ (brown marker and solid line) is almost linear. The relationship between $D_\mathrm{6.2}$ and $p_\mathrm{BA1}$ (orange marker and solid line) and that between $D_\mathrm{6.1}+D_\mathrm{6.2}$ and $p_\mathrm{BA1}$ (red marker and solid line) are almost linear in the region from 0 to 87.1\% of $p_\mathrm{BA1}$.

Table \ref{table:corr_inf_ba1_wave6} shows the respective sample correlation coefficient $\gamma$. The value of $\gamma$ for $D_\mathrm{6.1}+D_\mathrm{6.2}$ is 0.990, indicating the strongest correlation. In the region where $p_\mathrm{BA1}$ exceeds 87.1\%, $D_\mathrm{6.2}$ and $D_\mathrm{6.3}$ deviate from the straight line and increase rapidly.

% --- Table of sample correlation coefficient --- 
\begin{table} %[h] 
  \caption {The sample correlation coefficient $\gamma$ of the relationship between each of $D_\mathrm{6.1}, D_\mathrm{6.2}, D_\mathrm{6.1}+D_\mathrm{6.2}, D_\mathrm{6.3}$ and $p_\mathrm{BA1}$.\\} 
  \label {table:corr_inf_ba1_wave6} 
  \centering 
  \begin {tabular} {ccccc} 
	\hline 
	Wave & \multicolumn{1}{c}{$D_\mathrm{6.1}$} & \multicolumn{1}{c}{$D_\mathrm{6.2}$} & \multicolumn{1}{c}{$D_\mathrm{6.1}+D_\mathrm{6.2}$} & \multicolumn{1}{c}{$D_\mathrm{6.3}$}\\ 
	\hline \hline 
	$\gamma$ & 0.923 & 0.977 & 0.990 & 0.889 \\ 
	\hline
  \end {tabular} 
\end {table}

% --- subsection --- 
\subsection{The correlation between the BA.1 variant ratio and the infections among people in their 20s}

A characteristic increase in infections among people in their 20s can be seen in the above Monitoring Meeting materials\cite{monitoring84}. Table \ref{table:inf20s_wave6} shows the percentage $p_\mathrm{20s}$ of the total infections $D_\mathrm{20s}$ among people in their 20s to $D_\mathrm{all}$ of all people on $t_{1}=$17-Dec-2021, $t_{2}=$24-Dec-2021, $t_{3}=$31-Dec-2021, and $t_{4}=$07-Jan-2022. Days are expressed as medians.

% --- Table of sample correlation coefficient --- 
\begin{table} %[h] 
  \caption {The percentage $p_\mathrm{20s}$ of the total infections $D_\mathrm{20s}$ among people in their 20s to $D_\mathrm{all}$ of all people on $t_{1}=$17-Dec-2021, $t_{2}=$24-Dec-2021, $t_{3}=$31-Dec-2021, and $t_{4}=$07-Jan-2022.\\} 
  \label {table:inf20s_wave6} 
  \centering 
  \begin {tabular} {crrrr} 
	\hline 
	Day & \multicolumn{1}{c}{$t_{1}$} & \multicolumn{1}{c}{$t_{2}$} & \multicolumn{1}{c}{$t_{3}$} & \multicolumn{1}{c}{$t_{4}$}\\ 
	\hline \hline 
	$p_\mathrm{20s}$ & 18.8 & 25.9 & 28.0 & 40.5 \\ 
	\hline
  \end {tabular} 
\end {table}

\begin {figure} %[p] 
	\centering 
	\includegraphics [width=8.0cm] {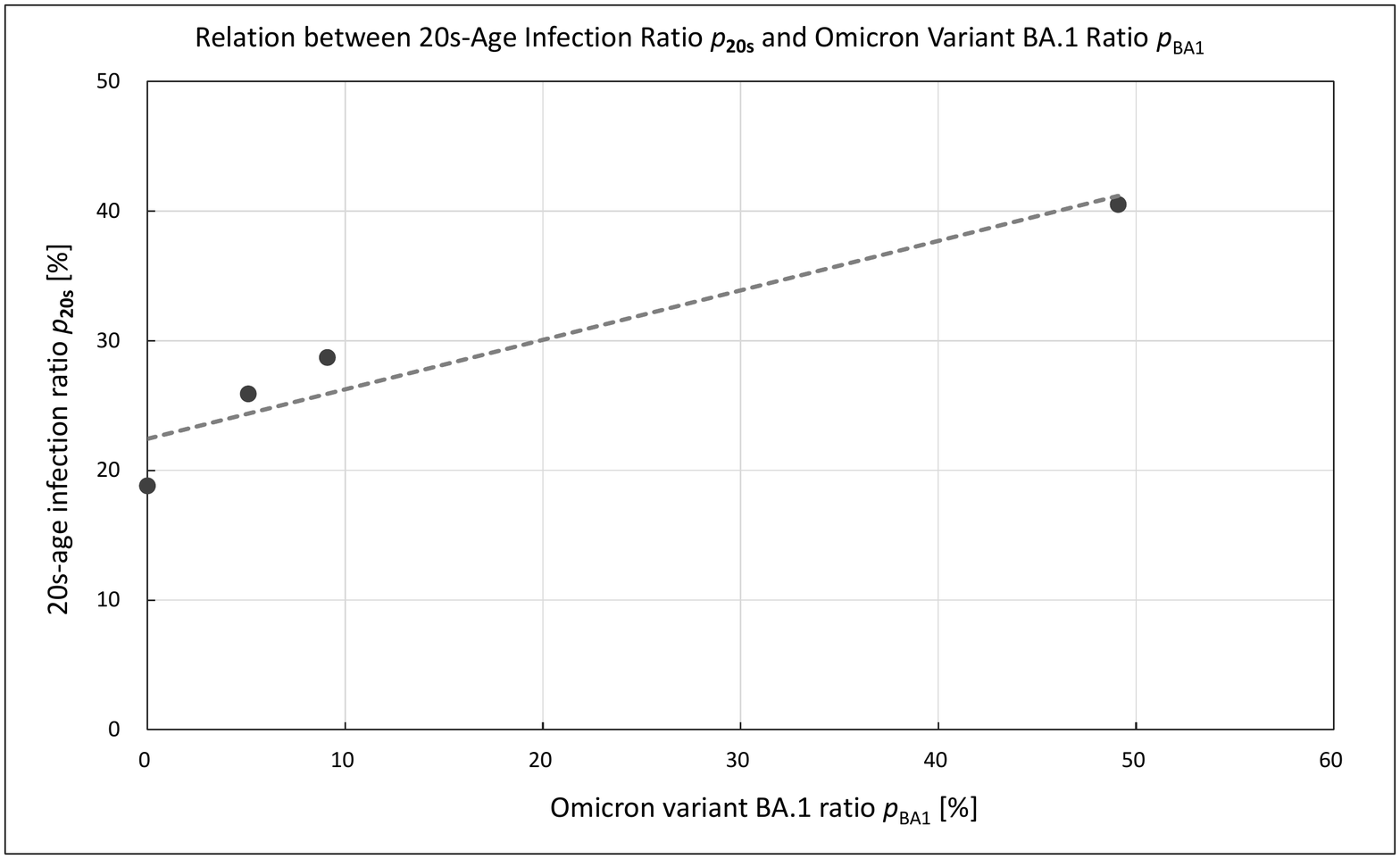}
	\caption {The relationship between  the 20s age infection ratio $p_\mathrm{20s}$ and $p_\mathrm{BA1}$.
	}
	\label {fig:ba1_inf20s_wave6} 
\end {figure}

Since this increase appears to correspond to an increase in $p_\mathrm{BA1}(t)$, Fig. \ref{fig:ba1_inf20s_wave6} shows the relationship between $p_\mathrm{20s}$ and $p_\mathrm{BA1}$. The data from 17-Dec-2021 to 07-Jan-2022 are almost linear, with a high sample correlation coefficient of 0.950.

% --- subsection --- 
\subsection{Factors behind outbreak of main component of the 6th wave}

During the 6th wave, two events occurred in which the main epidemic virus was replaced by a more infectious variant. 

As shown in Fig. \ref{fig:inf_ba1_wave6}, the infection cases with the Omicron variant were confirmed in Tokyo in mid-December 2021, when the 5th wave was converging. Since the Omicron variant was more infectious than the Delta variant, infections began to increase, and its ratio increased in proportion to the total number of new infections, forming relatively small waves 6.1 and 6.2 (see Fig. \ref{fig:allwaves_semilog}). This is the beginning of the first replacement. 

There were New Year holidays (31-Dec-2021 to 03-Jan-2022) during the 6.2th wave. In New Year holidays, as is the custom of the Japanese people, there is a great increase in the exchange of people. Table \ref{table:inf20s_wave6} shows that the number of new infections among people in their 20s increased significantly around the New Year holidays, which is thought to be the result of active interaction among people in their 20s.

As the number of infections increased significantly, it was newly regarded as the outbreak of the 6.3th wave (see Fig. \ref{fig:wave6}). When the 6.3th wave occurred (02-Jan-2022), the value of $p_\mathrm{BA1}(t)$ was 9.1\%, but by 20-Jan-2022, when the number of daily infections reached 50\% of its maximum, it was 97.1\%, reaching 99.6\% on 05-Feb-2022, about a month later.

From this process, it was thought that the coupling of increased human interaction due to the New Year holidays and the invasion of the new virus variant Omicron BA.1 caused the nucleation and domain growth of the Omicron variant in places and among people where nucleation was difficult to occur with the Delta variant, and then the outbreak of the dominant 6.3th wave occurred.

% --- subsection --- 
\subsection{Factors behind the outbreak of the three side-waves of the 6th wave}

After the dominant 6.3th wave, the second replacement of virus variant occurred.

\begin {figure} %[p] 
	\centering 
	\includegraphics [width=8.0cm] {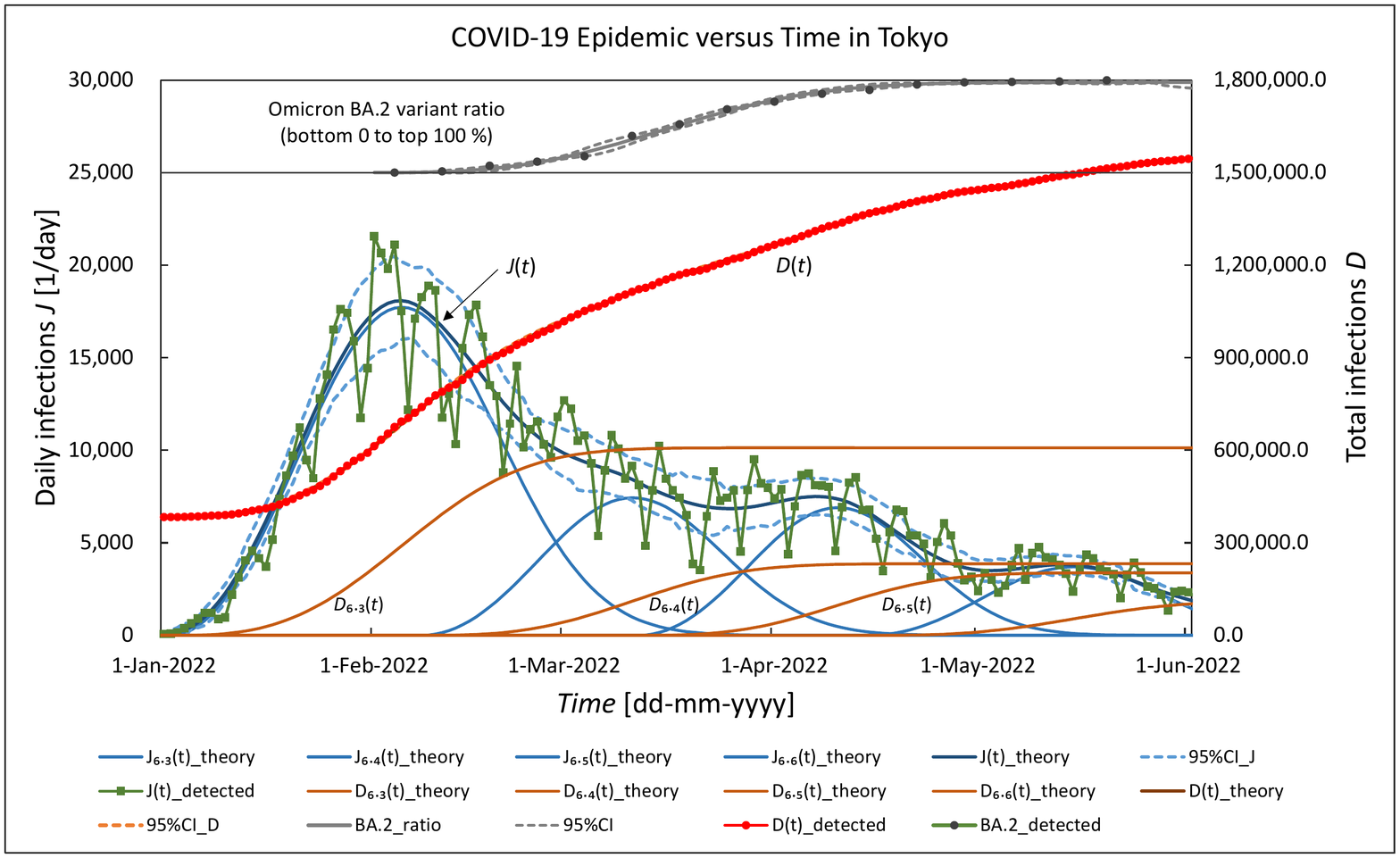}
	\caption {The characteristics of the 6th wave with the data of $p_\mathrm{BA2}(t)$. Symbols: refer to Fig. \ref{fig:wave6}.
	}
	\label {fig:inf_ba2_wave6} 
\end {figure}

Figure \ref{fig:inf_ba2_wave6} shows the characteristics of the 6th wave with the percentage $p_\mathrm{BA2}(t)$ of the suspected Omicron BA.2 variant\cite{voc_who} from the (96th) Tokyo Metropolitan Government Novel Coronavirus Infectious Diseases Monitoring Meeting materials\cite{monitoring96} inserted in the upper part.

\begin {figure} %[p] 
	\centering 
	\includegraphics [width=8.0cm] {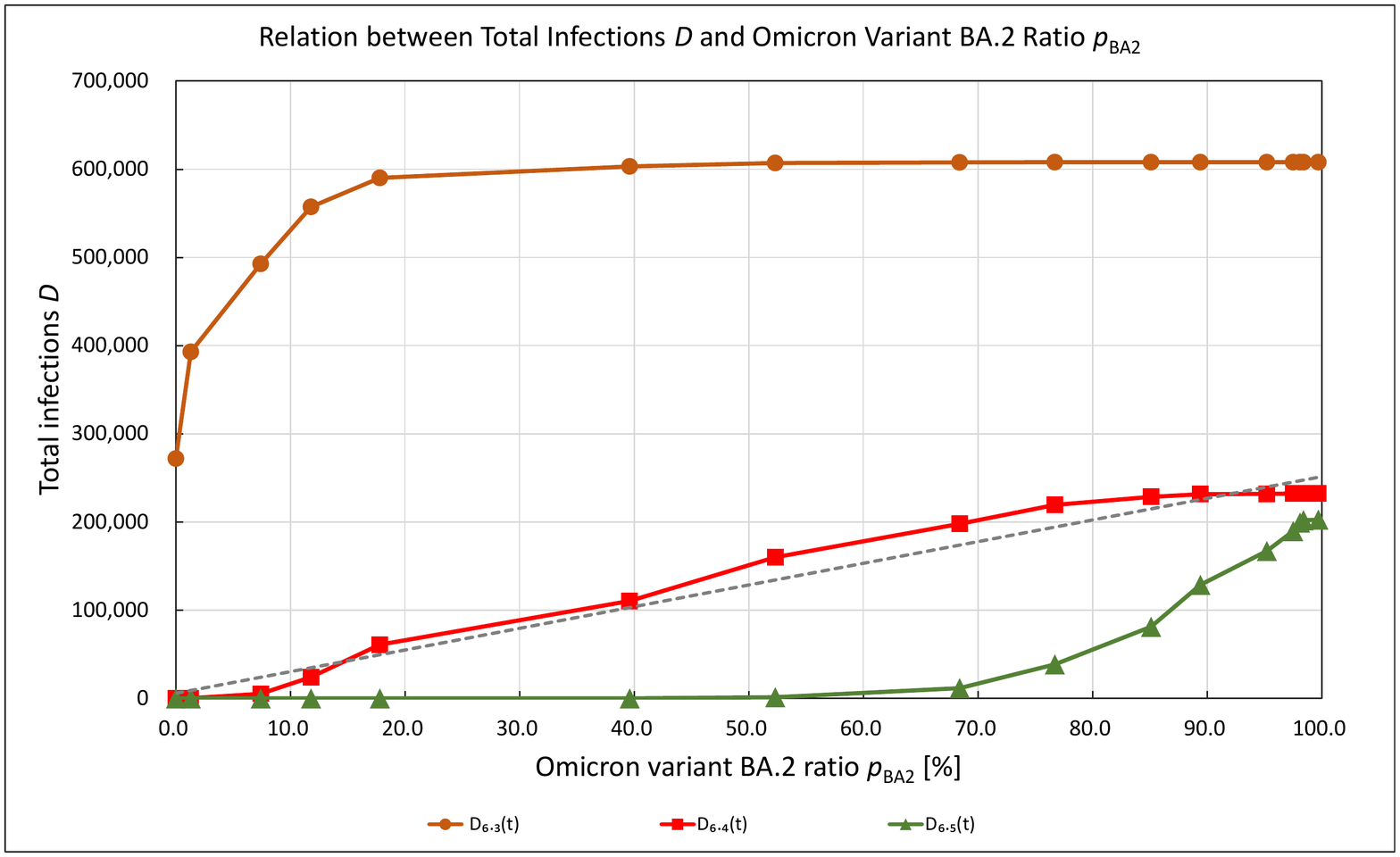}
	\caption {The relationship between $D_{\mathrm{6.3}}, D_{\mathrm{6.4}}, D_{\mathrm{6.5}}$ and $p_\mathrm{BA2}$.
	}
	\label {fig:corr_inf_ba2_wave6}
\end {figure}

In order to know the quantitative correlation, the relationship between $D_\mathrm{6.3}, D_\mathrm{6.4}, D_\mathrm{6.5}$ and $p_\mathrm{BA2}$ is shown in Fig. \ref{fig:corr_inf_ba2_wave6}, and sample correlation coefficient are shown in Table \ref{table:corr_inf_ba2_wave6}.
The relationship between $D_\mathrm{6.4}$ and $p_\mathrm{BA2}$ was almost linearly approximated, and the sample correlation coefficient was 0.986, indicating the strongest correlation.

% --- Table of sample correlation coefficient --- 
\begin{table} %[h] 
  \caption {The sample correlation coefficient $\gamma$ of the relationship between each of $D_\mathrm{6.3}, D_\mathrm{6.4}, D_\mathrm{6.5}$ and $p_\mathrm{BA2}$.\\} 
  \label {table:corr_inf_ba2_wave6}
  \centering 
  \begin {tabular} {crrrr} 
	\hline 
	Wave & \multicolumn{1}{c}{$D_\mathrm{6.3}$} & \multicolumn{1}{c}{$D_\mathrm{6.4}$} & \multicolumn{1}{c}{$D_{\mathrm{6.5}}$}\\ 
	\hline \hline 
	$\gamma$ & 0.717 & 0.986 & 0.853  \\ 
	\hline
  \end {tabular} 
\end {table}

As shown in Fig. \ref{fig:wave6}, the 6.4th wave occurred around the February holidays (11 to 13-Feb-2022). The value of $p_\mathrm{BA2}(t)$ increased in proportion to the total infections of the 6.4th wave, and exceeded 95\% around 20-Apr-2022, about two months later.
The 6.5th wave occurred around the consecutive holidays in March (19 to 21-Mar-2022), and the 6.6th wave occurred around the consecutive holidays in May (29-Apr to 05-May-2022).
 Although these consecutive holiday effects were not as large as the New Year holyday effect, they were thought to have reduced the decreasing trend of the 6th wave and prevented it from converging for a long period of time coupled with the new virus variant BA.2 invasion.

% --- Discussions of the 7th wave --- 
\section{Waveform analysis of the COVID-19 7th infection in Tokyo} 
% --- subsection The 7th wave in Tokyo --- 
\subsection{Simulation characteristics of the 7th wave} 

\begin {figure} %[p] 
	\centering 
	\includegraphics [width=8.0cm] {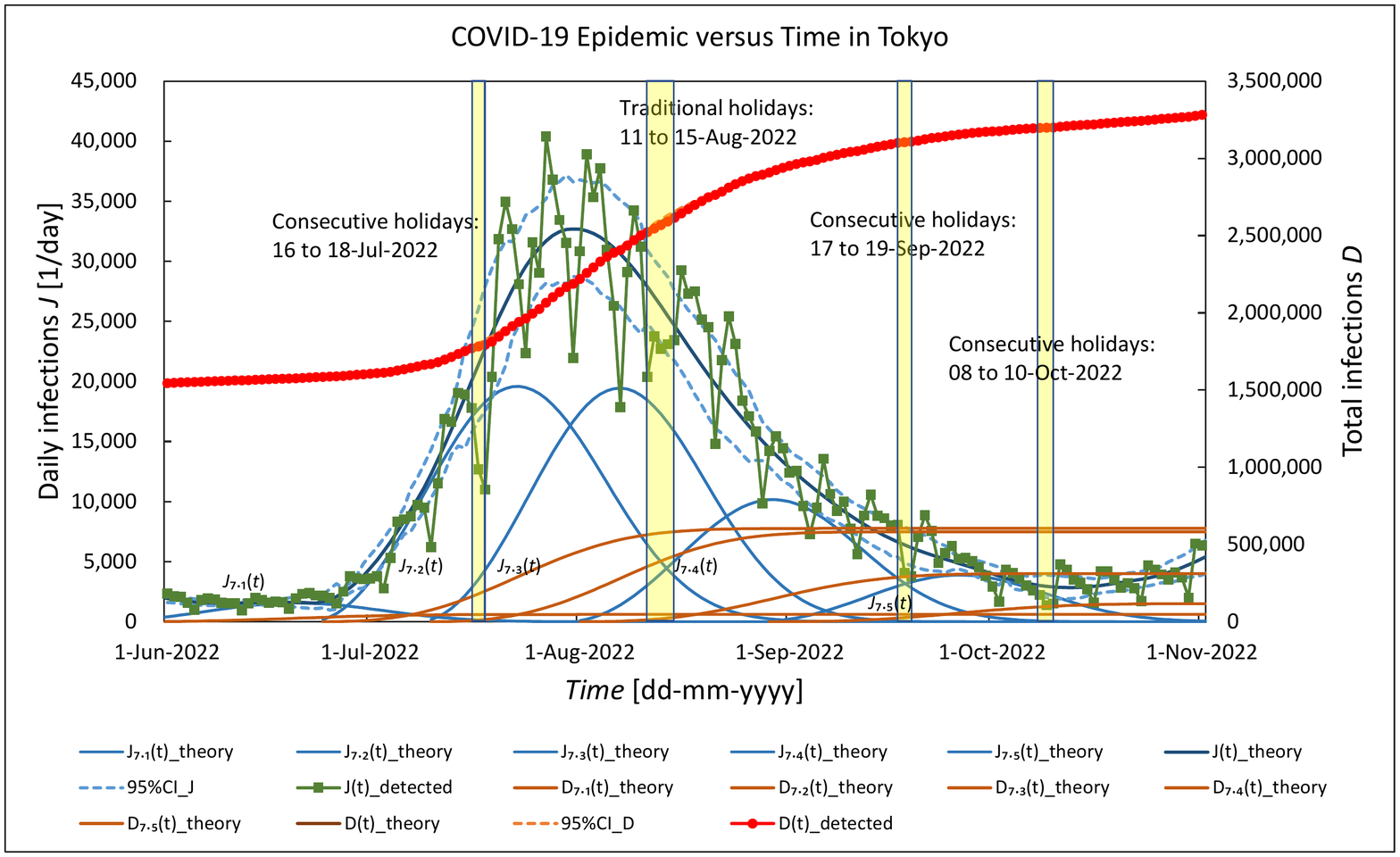}
	\caption {The detected and the simulated values of $J(t)$ and $D(t)$ of the 7th wave in Tokyo from June 1, 2022 to November 1, 2022. Symbols: refer to Fig. \ref{fig:wave6}.
	}
	\label {fig:wave7} 
\end {figure}

Figure \ref{fig:wave7} shows the detected and the simulated values of $J(t)$ and $D(t)$ for the 7th wave in Tokyo from June 1, 2022 to November 1, 2022. The meaning of the markers and each line is the same as that in Fig. \ref{fig:wave6}.

% --- subsection --- 
\subsection{The correlation between the 7th wave and the BA.5 variant ratio}

Shortly before the start of the 7th wave, the BA.2 variant was replaced by the more infectious BA.5 variant\cite{voc_who}.  

\begin {figure} %[p] 
	\centering 
	\includegraphics [width=8.0cm] {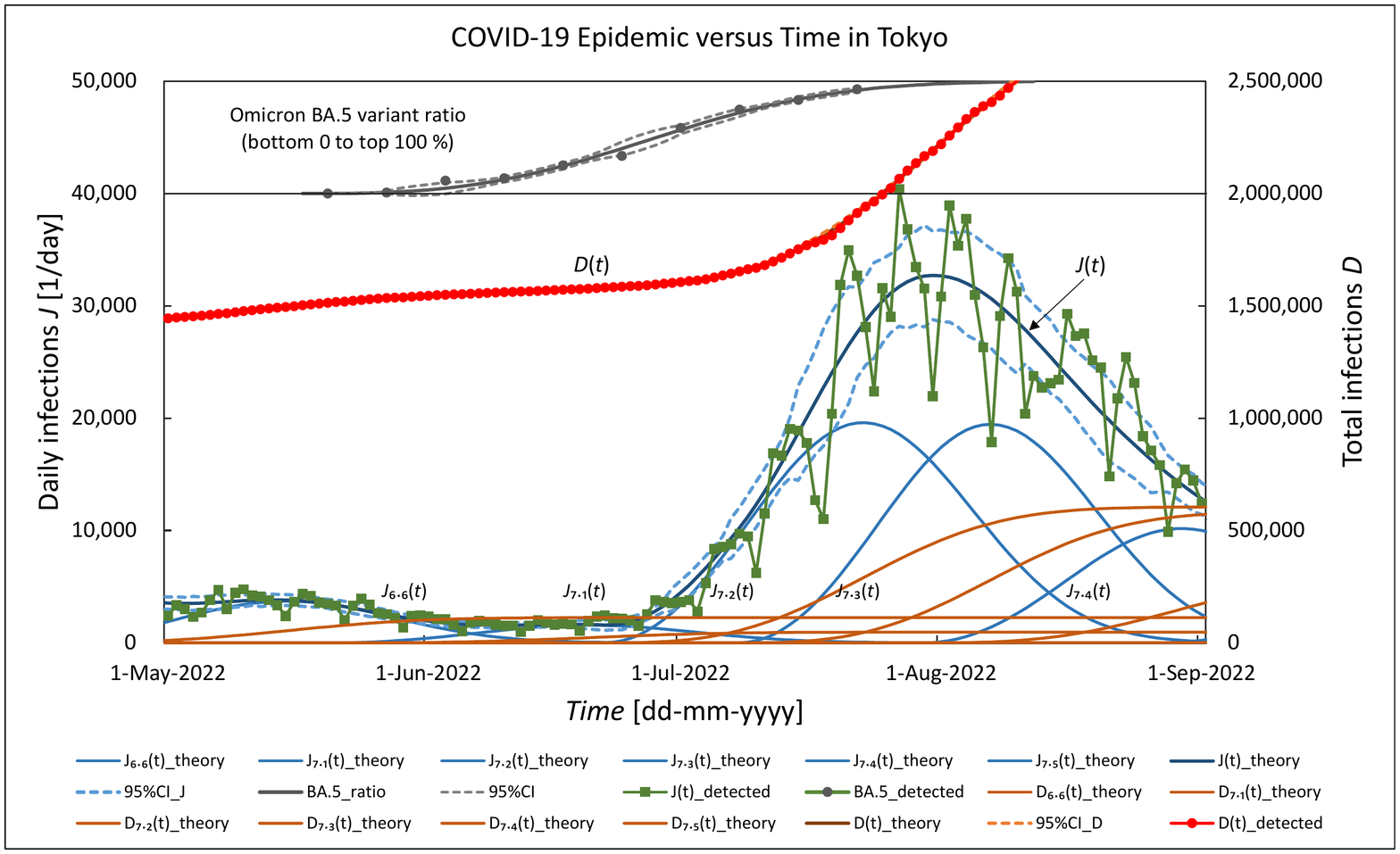}
	\caption {The characteristics of the 7th wave with the data of $p_\mathrm{BA5}(t)$. Symbols: refer to Fig. \ref{fig:wave6}.
	}
	\label {fig:inf_ba5_wave7} 
\end {figure}

Figure \ref{fig:inf_ba5_wave7} shows the characteristics of the 7th wave with the percentage $p_\mathrm{BA5}(t)$ of the suspected Omicron BA.5 variant from the (96th) Monitoring Meeting materials\cite{monitoring96} inserted in the upper part.
The 7.1th wave occurred around 20-May-2022. The value of $p_\mathrm{BA5}(t)$ increased almost in proportion to the total infections during the 7.1th wave, reaching 92.8\% around 22-Jul-2022, about two months later. There are no consecutive holidays in June, so there is no consecutive holiday effect.

\begin {figure} %[p] 
	\centering 
	\includegraphics [width=8.0cm] {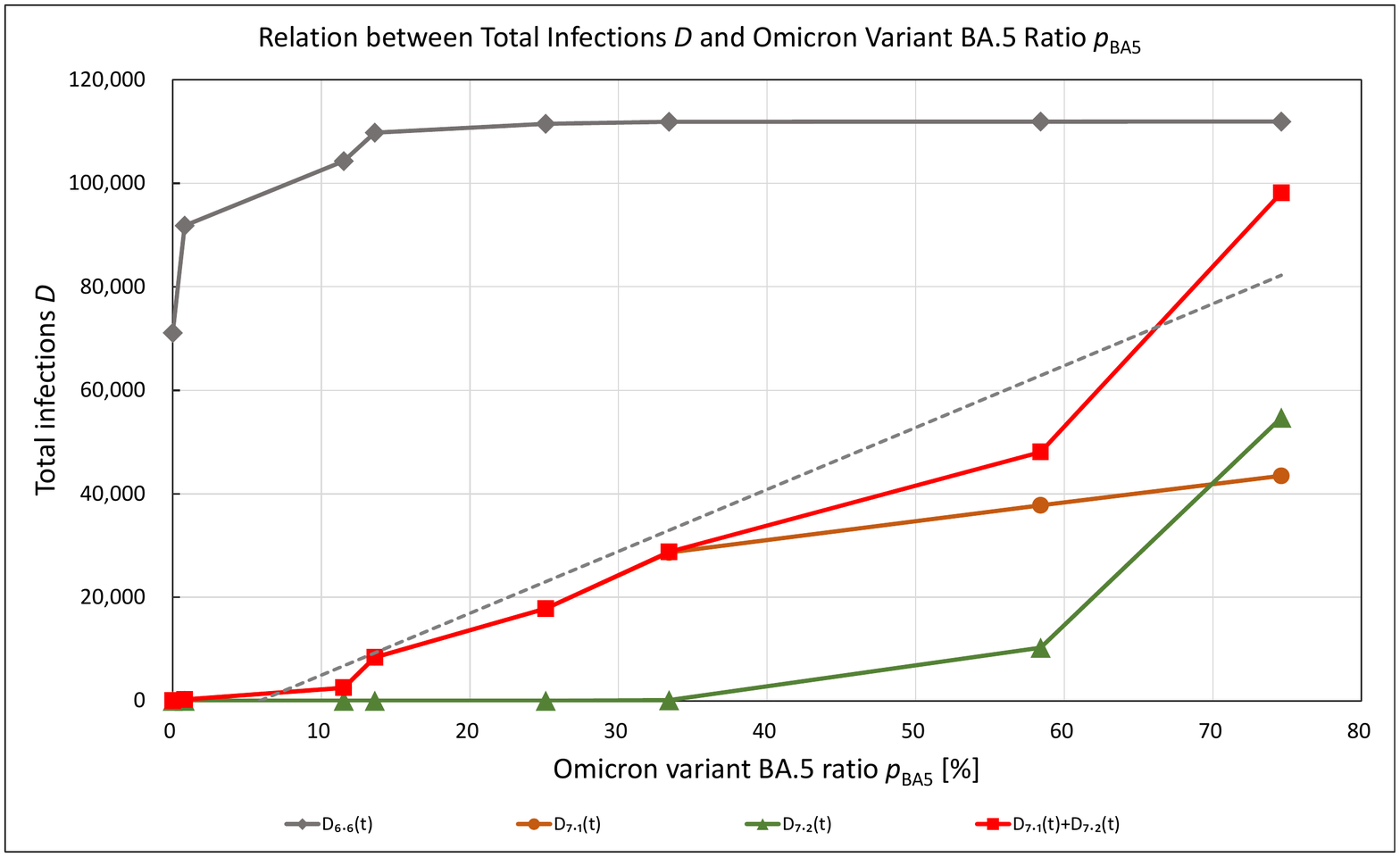}
	\caption {The relationship between $D_\mathrm{6.6}, D_\mathrm{7.1}, D_\mathrm{7.2}, D_\mathrm{7.1}+D_\mathrm{7.2}$ and $p_\mathrm{BA5}$. 
	}
	\label {fig:corr_inf_ba5_wave7} 
\end {figure}

Figure \ref{fig:corr_inf_ba5_wave7} shows the relationship between $D_\mathrm{6.6}, D_\mathrm{7.1}, D_\mathrm{7.2}, D_\mathrm{7.1}+D_\mathrm{7.2}$ and $p_\mathrm{BA5}$ to quantitatively understand the virus mutation effect. The relationship between $D_\mathrm{7.1}$ and $p_\mathrm{BA5}$, and between $D_\mathrm{7.1}+D_\mathrm{7.2}$ and $p_\mathrm{BA5}$ (brown and red markers and solid lines, respectively) are almost linear. The relationship between $D_\mathrm{7.2}$ and $p_\mathrm{BA5}$ (dark green marker and solid line) is almost linear in the region from 0 to about 60\% of $p_\mathrm{BA5}$.

Table \ref{table:corr_inf_ba5_wave7} shows the respective sample correlation coefficient $\gamma$. The value of $\gamma$ for $D_\mathrm{7.1}$ is 0.977, indicating the strongest correlation. In the region where $p_\mathrm{BA5}$ exceeds 60\%, $D_\mathrm{7.2}$ deviates from the straight line and increases rapidly.

% --- Table of sample correlation coefficient --- 
\begin{table} %[h] 
  \caption {The sample correlation coefficient $\gamma$ of the relationship between each of $D_\mathrm{6.6}, D_\mathrm{7.1}, D_\mathrm{7.2}, D_\mathrm{7.1}+D_\mathrm{7.2}$ and $p_\mathrm{BA5}$.\\}
  \label {table:corr_inf_ba5_wave7} 
  \centering 
  \begin {tabular} {ccccc} 
	\hline 
	Wave & \multicolumn{1}{c}{$D_\mathrm{6.6}$} & \multicolumn{1}{c}{$D_\mathrm{7.1}$} & \multicolumn{1}{c}{$D_\mathrm{7.2}$} & \multicolumn{1}{c}{$D_\mathrm{7.1}+D_\mathrm{7.2}$}\\ 
	\hline \hline 
	$\gamma$ & 0.645 & 0.977 & 0.803 & 0.960 \\ 
	\hline
  \end {tabular} 
\end {table}

% --- subsection --- 
\subsection{The correlation between the 7th wave and the daily maximum temperature}

The daily maximum temperature $T_\mathrm{max}(t)$ was approximately 20$^\circ$C in early June, but exceeded 35$^\circ$C in late June and reached 37.0$^\circ$C on July 1st. When the temperature rises, air conditioners will be used indoors, and there will be an increase in the tendency for families to gather in one room without wearing masks, especially at home.
 
\begin {figure} %[p] 
	\centering 
	\includegraphics [width=8.0cm] {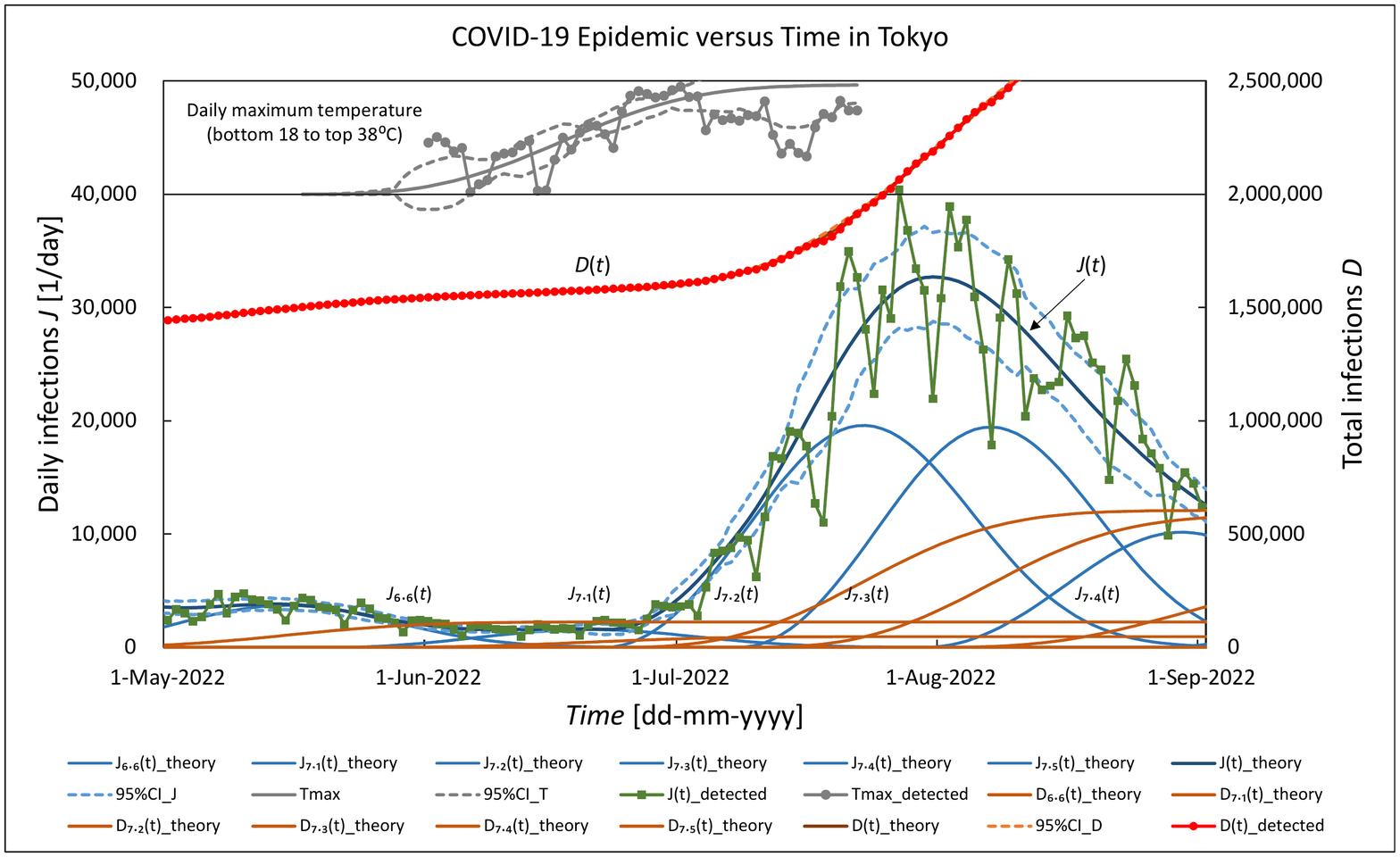} 
	\caption {The characteristics of the 7th wave with the daily maximum temperature. Symbols: refer to Fig. \ref{fig:wave6}.
	}
	\label {fig:inf_tmax_wave7}
\end {figure}

Figure \ref{fig:inf_tmax_wave7} shows the characteristics of the 7th wave with $T_\mathrm{max}(t)$\cite{meteorological} inserted above. 
Comparing Fig. \ref{fig:inf_ba5_wave7} and Fig. \ref{fig:inf_tmax_wave7}, it is expected that the characteristics of $p_\mathrm{BA5}(t)$ and $T_\mathrm{max}(t)$ are highly correlated.

\begin {figure} %[p] 
	\centering 
	\includegraphics [width=8.0cm] {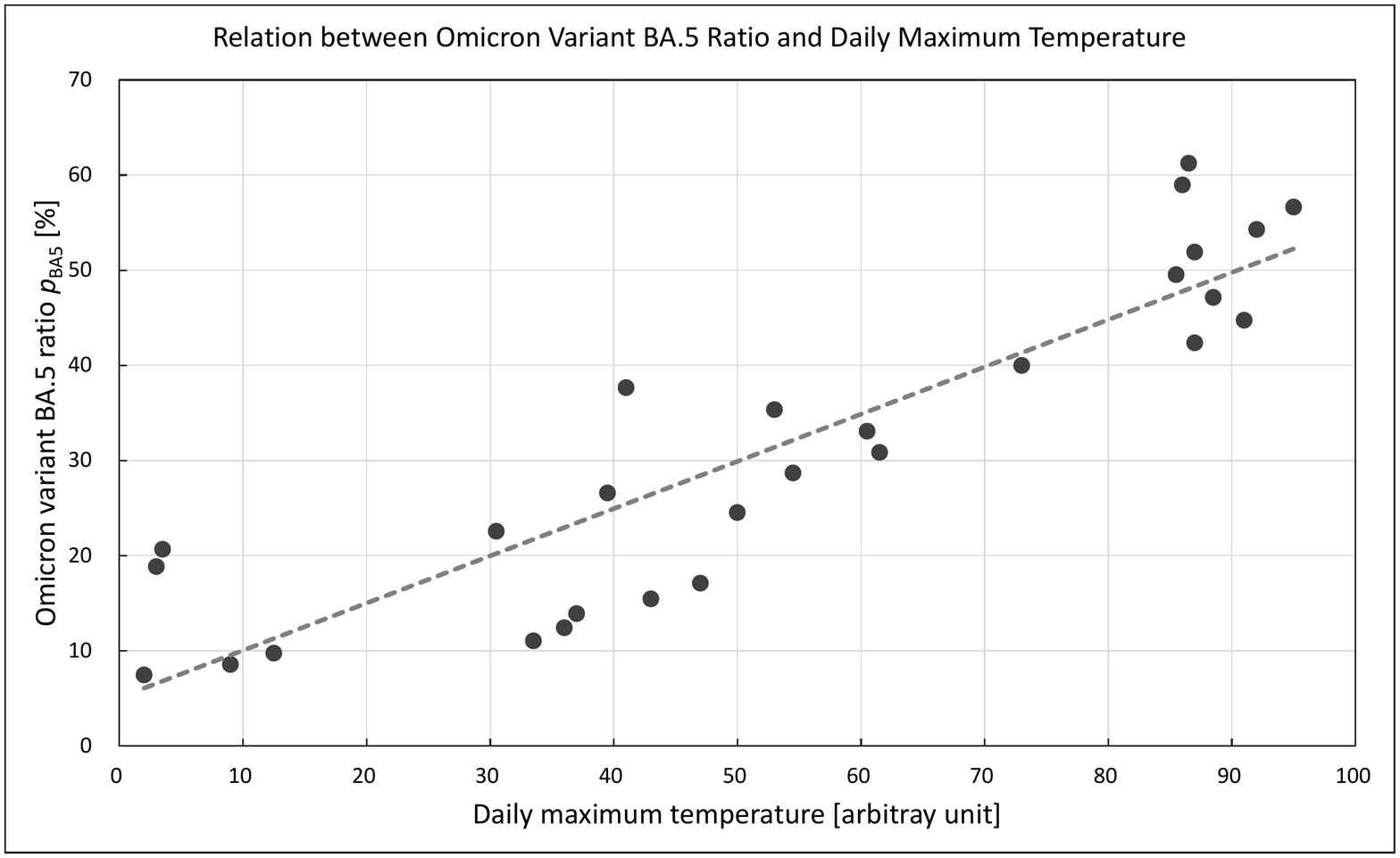}
	\caption {The relationship between $p_\mathrm{BA5}$ and ($T_\mathrm{max} - 18.0$).
	}
	\label {fig:corr_ba5_tmax_wave7}
\end {figure}

The relationship between $p_\mathrm{BA5}$ and ($T_\mathrm{max} - 18.0$) is shown in Fig. \ref{fig:corr_ba5_tmax_wave7}. Although the temperature fluctuates, the value of $\gamma$ is 0.922, showing a strong correlation.

% --- subsection --- 
\subsection{High infection rate among young people}

Table \ref{table:inf_agegroup} shows the percentage of new infections by age group\cite{monitoring96}. The combined infection rate for under-10s and teenagers is about 22\%, higher than the highest other age group, about 20\%, for those in their 20s.

% --- Table of three paremeters --- 
\begin{table} [h] 
  \caption {The percentage of new infections by age group on the period of (a) 19-Jul-2022 to 25-Jul-2022 and (b) 26-Jul-2022 to 01-Aug-2022. \\}
  \label {table:inf_agegroup}
  \centering 
  \begin {tabular} {crrrrrrrr} 
	\hline 
	Age & \multicolumn{1}{c}{$<$10s} & \multicolumn{1}{c}{10s} & \multicolumn{1}{c}{20s} & \multicolumn{1}{c}{30s} & \multicolumn{1}{c}{40s} & \multicolumn{1}{c}{50s}\\ 
	\hline \hline 
	(a) & 10.3 & 13.1 & 20.1 & 16.9 & 15.5 & 11.8\\ 
	(b) & 9.8 & 11.0 & 19.3 & 17.0 & 17.2 & 12.8\\
	\hline
  \end {tabular} 
\end {table}

% --- subsection --- 
\subsection{High rate of infection from cohabitants}

About 75\% of newly infected people have unknown contact history. Table \ref{table:inf_routegroup} shows the percentage of newly infected people by infection route whose infection route was clear\cite{monitoring96}. Infection from people who live together is the most common, about 68\%.

% --- Table of three paremeters --- 
\begin{table} %[h] 
  \caption {The percentage of newly infected people by infection route, (1) living together, (2) facilities, (3) workplace, (4) dinning together, (5) dinning together with entertainment, and (6) others on the period of (a) 19-Jul-2022 to 25-Jul-2022 and (b) 26-Jul-2022 to 01-Aug-2022. \\}
  \label {table:inf_routegroup}
  \centering 
  \begin {tabular} {crrrrrr} 
	\hline 
	Route & \multicolumn{1}{c}{(1)} & \multicolumn{1}{c}{(2)} & \multicolumn{1}{c}{(3)} & \multicolumn{1}{c}{(4)} & \multicolumn{1}{c}{(5)} & \multicolumn{1}{c}{(6)} \\ 
	\hline \hline 
	(a) & 68.7 & 15.2 & 6.6 & 2.2 & 0.0 & 7.4 \\ 
	(b) & 67.9 & 14.3 & 7.4 & 1.7 & 0.0 & 8.7 \\
	\hline
  \end {tabular} 
\end {table}

% --- subsection --- 
\subsection{Low vaccination coverage among younger age groups}

Table \ref{table:vaccin_agegroup} shows the percentage $p_{v}$ of the third vaccination by age group in Tokyo (updated on August 8, 2022)\cite{booster_by_age}. In the 12-19 age group, it is the lowest at 37.4\%. Vaccination for the 5-11 age group is said to be carried out, but no data is shown.

% --- Table of three paremeters --- 
\begin{table} %[h] 
  \caption {The percentage $p_{v}$ of the vaccination by age group.\\} 
  \label {table:vaccin_agegroup}
  \centering 
  \begin {tabular} {crrrrrrrr} 
	\hline 
	Age & \multicolumn{1}{c}{12-19} & \multicolumn{1}{c}{20s} & \multicolumn{1}{c}{30s} & \multicolumn{1}{c}{40s} & \multicolumn{1}{c}{50s} & \multicolumn{1}{c}{60s} \\ 
	\hline \hline 
	$p_{v}$ & 37.4 & 47.5 & 54.7 & 62.6 & 79.0 & 86.8 \\ 
	\hline
  \end {tabular} 
\end {table}

% --- subsection --- 
\subsection{Complex factors behind the outbreak of the 7th wave}

The coupling of multiple factors indicated above and the invasion of the new virus variant BA.5 was thought to lead to the 7th wave outbreak in the following procedure.

First, the nucleation of the BA.5 variant occurred predominantly over the BA.2 variant in young people. Infection spread among people who were in daily contact with those nuclei. This infection formed a large initial 7.2th wave. Furthermore, the nucleation of the BA.5 variant coupled with the consecutive holidays in July (16 to 18-Jul-2022, refer to Fig. \ref{fig:wave7}) spread the infection to middle-aged and elderly people in addition to young people, causing the equally large 7.3th wave. The superposition of these two waves formed the main component of the 7th wave, the largest ever (see Fig. \ref{fig:wave7}).

% --- Discussions --- 
\section {Discussions}
% --- subsection --- 
\subsection {Dense nucleation followed by a near growth model}

The present analysis is based on the random nucleation and subsequent two dimensional  linear domain growth model\cite{sim_covid19_1to5_japan,covid19_tokyo}.
The values of $K$ in Table \ref{table:three_parameters} are almost constant, about 0.10 for the 5.1th to the 6.3th waves (A group), and about 0.13 for the 6.4th to the 7.5th waves (B group). 

\begin {figure} %[p] 
	\centering 
	\includegraphics [width=8.0cm] {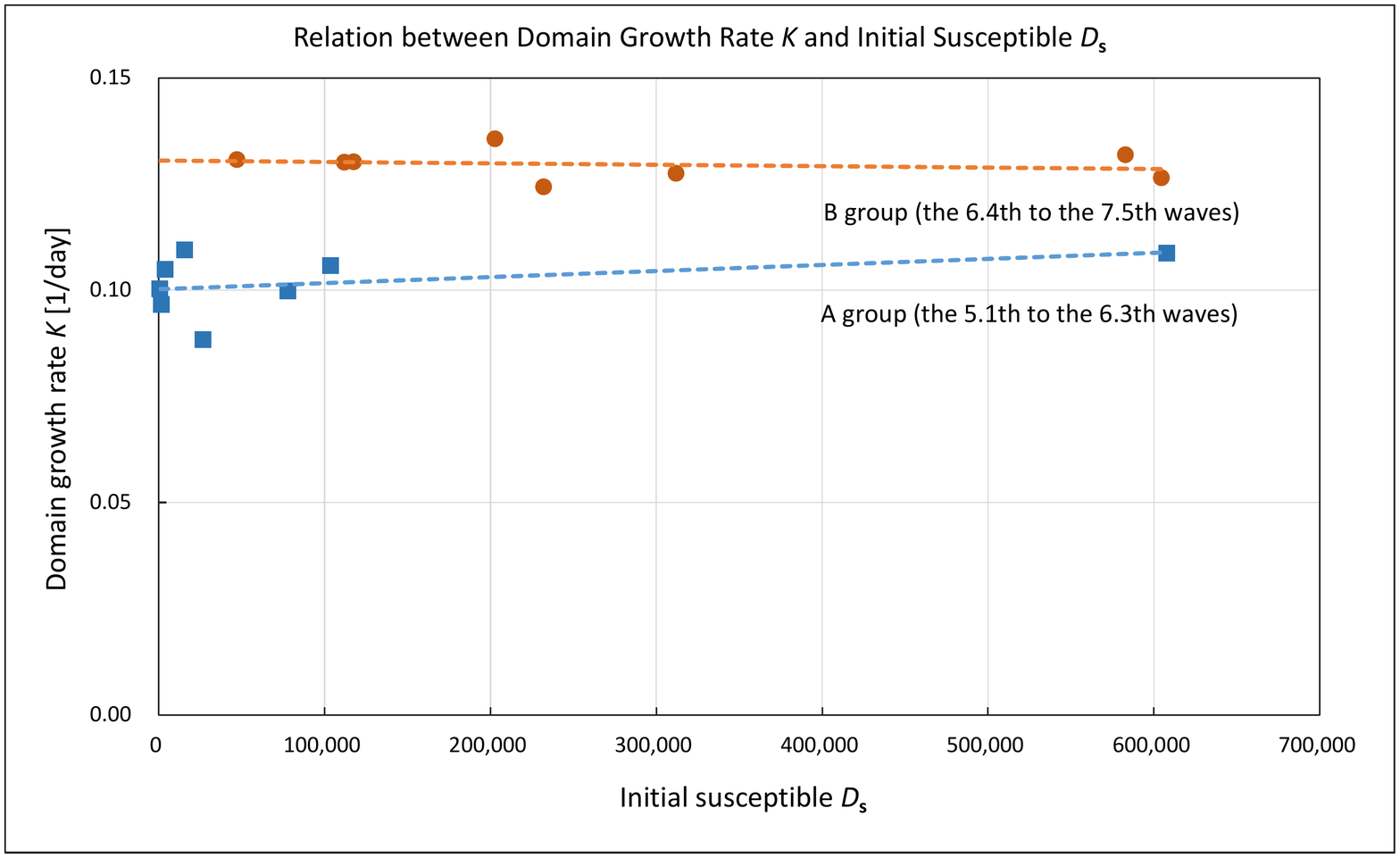}
	\caption {The relationship between $K$ and $D_\mathrm{s}$ for A and B groups.
	}
	\label {fig:k_ds_wave} 
\end {figure}

The relationship between $K$ and $D_\mathrm{s}$ is shown in Fig. \ref{fig:k_ds_wave} for A and B groups. 
The value of $K$ tends to increase slightly with increasing $D_\mathrm{s}$ for A group. The value of $K$ is less than 1.3 times greater even when the value of $D_\mathrm{s}$ increases by a factor of about 900 (refer to Table \ref{table:three_parameters}). For B group, the value of $K$ is almost constant without depending on $D_\mathrm{s}$ (refer to Table \ref{table:three_parameters}).

\begin {figure} [p] 
	\centering 
	\includegraphics [width=8.0cm] {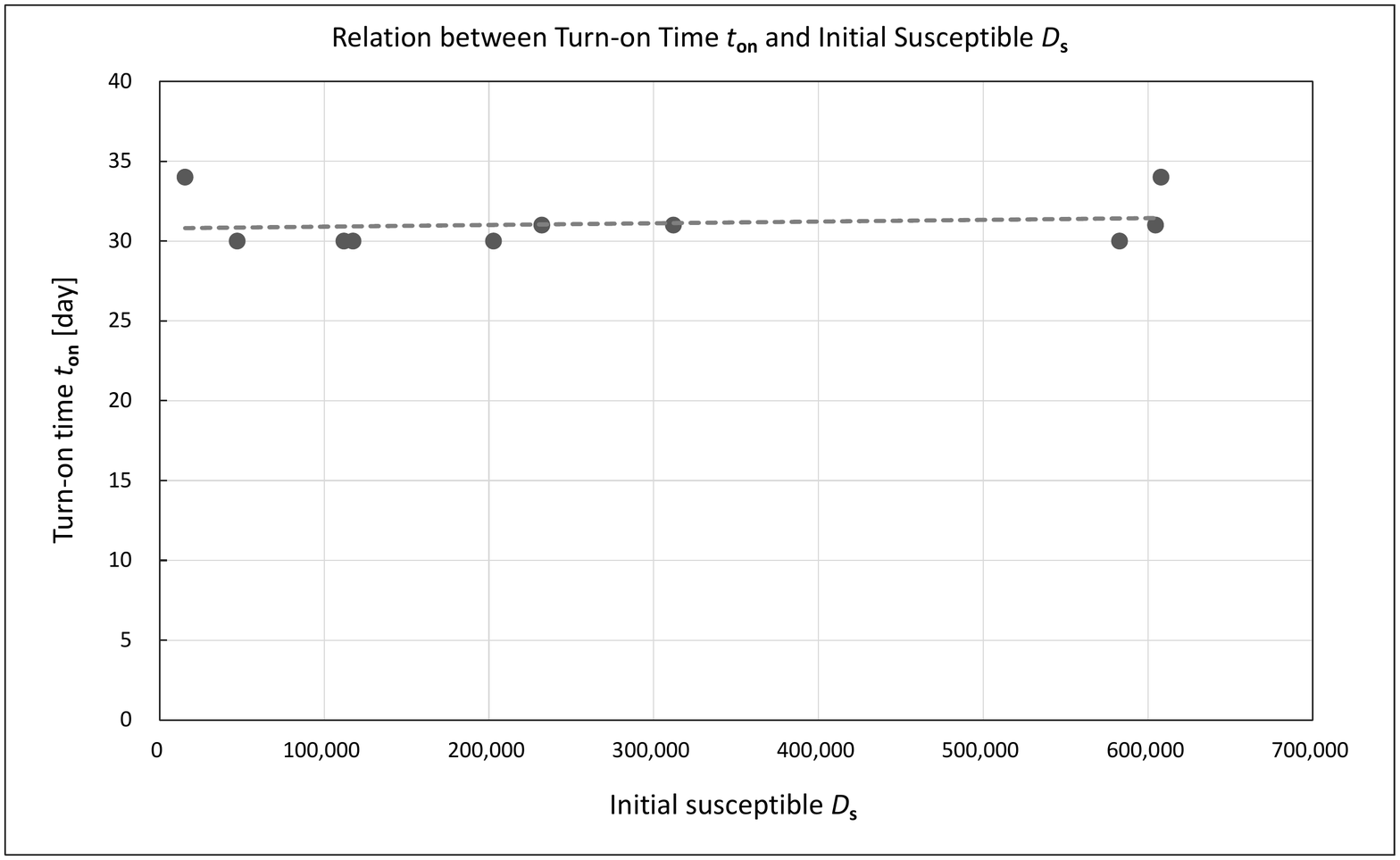}
	\caption {The relationship between $t_\mathrm{on}$ and $D_\mathrm{s}$ from the 6.2th and the 7.5th waves.
	}
	\label {fig:ton_ds_wave6_7} 
\end {figure}

Figure \ref{fig:ton_ds_wave6_7} shows the relationship between the turn-on time $t_\mathrm{on}$ and $D_\mathrm{s}$ from the 6.2th and the 7.5th waves. Even if the value of $D_\mathrm{s}$ increases by about 40 times, the value of $t_\mathrm{on}$ changes by 1.1 times at most and is about 31 days (refer to Table \ref{table:three_parameters}).

Based on these characteristics, a model in which a few nuclei are generated and grow over a wide area is denied, and a model in which a lot of nuclei are generated randomly and subsequent domain growth reaches the growth limit in a relatively short period of time is deduced. 

Large waves are mainly caused by nucleation in many places, resulting in large $D_\mathrm{s}$. Although there is a tendency for $D_\mathrm{s}$ to increase as the growth rate increases, its contribution is small. 

Data on vaccination coverage and number of new PCR positives during the 7th wave are also consistent with this model. Table \ref{table:vacc_inf_agegroup} shows $A$: population by age group\cite{vacc_by_age}, $B$: the 3rd vaccination percentage\cite{booster_by_age}, $C = A \times (1 – B)$: number of unvaccinated people, $E$: number of new PCR positives during the 7th wave period (29-Jun-2022 to 02-Aug-2022)\cite{positives_by_age}, and $F = E / C$ [\%]: the percentage of the number of new PCR positives to the number of unvaccinated people.

% --- Table of three paremeters --- 
\begin{table} [h] 
  \caption {The third vaccination percentage and number of new PCR positives by age group. \\
  $A$: Population, $B$: Vaccination percentage, $C = A \times (1 - B)$, $E$: New PCR positives from 29-Jun-2022 to 02-Aug-2022, $F = E / C$ [\%]. \\
}
  \label {table:vacc_inf_agegroup}
  \centering 
  \begin {tabular} {crrrrr} 
	\hline 
	Age & \multicolumn{1}{c}{12-19} & \multicolumn{1}{c}{20s} & \multicolumn{1}{c}{30s} \\ 
	\hline \hline 
	$A$ & 9,010,292 & 12,819,569 & 14,372,705 \\ 
	$B$ & 35.6 & 48.6 & 52.3 \\
	$C$ & 5,802,628 & 6,589,258 & 6,855,780 \\
	$E$ & 537,335 & 562,450 & 549,759 \\
	$F$& 9.3 & 8.5 &8.0 \\
	\hline
	Age & \multicolumn{1}{c}{40s} & \multicolumn{1}{c}{50s}  \\ 
	\hline \hline 
	$A$ & 18,424,463 & 16,810,584 \\ 
	$B$ & 60.9 &78.1 \\
	$C$ & 7,203,965 & 3,681,518 \\
	$E$ & 563,307 & 389,904 \\
	$F$ & 7.8 &10.6 \\
	\hline
  \end {tabular} 
\end {table}

The number of people who have not been vaccinated for the third time, $C$, is the number of people who do not have sufficient antibodies against the BA.2 and the BA.5 variants that were the main cause of the 7th wave, so it is the number of people who can become newly PCR positive. The value of $E$ is less than 10\% ($F = E / C$) of the population with insufficient antibodies in the age group under 40. 

These data favor the dense nucleation followed by a near growth model, where a lot of nuclei are generated randomly at places where people gather, such as homes, facilities, working areas, and various gathering places. However, the domain growth is limited within the proximity of the nucleation.

% --- subsection --- 
\subsection {A quantity to represent the infectivity of the virus}

\begin {figure} %[p] 
	\centering 
	\includegraphics [width=8.0cm] {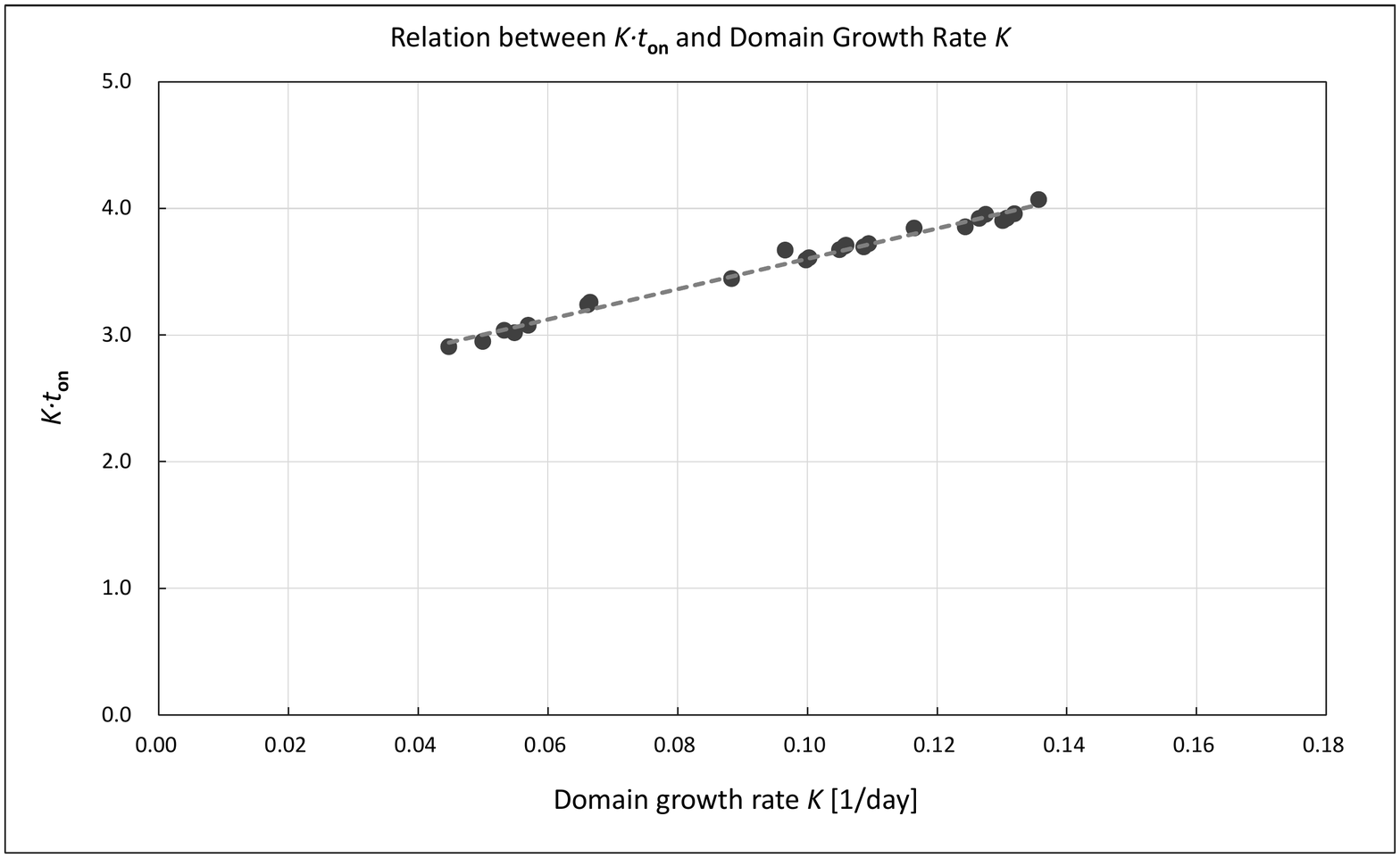}
	\caption {The relationship between the quantity $K \cdot t_\mathrm{on}$ and $K$ for all waves from the 1.1th to the 7.5th waves.
	}
	\label {fig:kton_k_allwaves} 
\end {figure}

Figure \ref{fig:kton_k_allwaves} shows the relationship between the quantity $K \cdot t_\mathrm{on}$ and $K$ for all waves from the 1.1th to the 7.5th waves. The value of $K \cdot t_\mathrm{on}$ increases almost linearly as the value of $K$ increases, and the deviation is small. If the unit of the growth rate is $number/day$ instead of $distance/time$, $K \cdot t_\mathrm{on}$ represents the number of persons undergoing one-dimensional linear growth in the turn-on time $t_\mathrm{on}$ from one generated nucleus. The average value for each group of this quantity is about 3.2 for the 1.1th to the 4.2th waves, about 3.6 for the 5.1th to the 6.3 waves, and about 4.0 for the 6.4th to the 7.5th waves. Since these values are almost independent of the magnitude of the infection wave, they are thought to represent the infectivity of the virus and are likely quantities related to the basic reproduction number.

% --- Conclusion --- 
\section {Conclusion}

The 6th and the 7th waves of COVID-19 in Tokyo were analyzed and well simulated by using the Avrami equation.
As a result, the factors that caused the 6th wave to last the longest and the 7th wave to be the largest were elucidated.

The main component of the 6th wave was formed by the coupling of increased human interaction due to the New Year holidays and the invasion of the new virus variant Omicron BA.1. After that, side waves were formed by the coupling of the invasion of the new virus variant Omicron BA.2 and the human interaction in the consecutive holidays in February, March, and May. These side waves caused the 6th wave not to converge for a long time.

The outbreak of the main component of the 7th wave occurred by the coupling of the invasion of the new virus variant Omicron BA.5 and multiple social factors, followed by human interaction during the July holidays.

Based on the results that the domain growth rate $K$ and the infection rise time $t_\mathrm{on}$ were almost independent of the initial susceptible $D_\mathrm{s}$, dense nucleation followed by a near growth model was deduced.

The quantity $K \cdot t_\mathrm{on}$ was considered to represent the infectivity of the virus.

\end{document}